\documentclass[11pt]{article}
\usepackage{moriond,epsfig}

\def\d{\rm d}

\def\as{\alpha_s}
\def\beq{\begin{equation}}
\def\eeq{\end{equation}}
\def\bea{\begin{eqnarray}}
\def\eea{\end{eqnarray}}
\begin{document}
\vspace*{4cm}
\title{Precision predictions for $Z^\prime$ production at the LHC}

\author{ Benjamin Fuks }

\address{Physikalisches Institut, Albert-Ludwigs-Universit\"at Freiburg, \\
Hermann-Herder-Stra\ss{}e 3, D-79106 Freiburg i.Br., Germany}

\maketitle\abstracts{
We present precision calculations of the transverse-momentum spectrum, the
invariant-mass distribution and the total cross section for $Z^\prime$
production at hadron colliders. We implement joint resummation at
the next-to-leading logarithmic accuracy and consistently match
the obtained result with the pure perturbative result at the first
order in the strong coupling constant. We confront our numerical predictions
with the Monte Carlo generator PYTHIA and with a new implementation of extra
neutral gauge bosons in the Monte Carlo generator MC@NLO. The impact of scale
dependence is also studied.}

%%%%%%%%%%%%%% Begin Section 1 %%%%%%%%%%%%%%%%%%%%%%%%%%%%%%%%%%%%%%%%%

\section{Introduction} \label{sec:1}

Despite its phenomenological success, the Standard Model (SM) of
particle physics is believed to suffer from a variety of conceptual
deficiencies. In particular, it provides no fundamental motivation why the
strong and electroweak interactions should be described by three different
gauge groups. Grand Unified Theories allow for a unification of these groups
which is broken to the SM at higher scale. This
yields one or several extra neutral gauge $Z^\prime$-bosons exhibiting the
existence of additional $U(1)$ symmetries \cite{Langacker:2008yv}. If the
$Z^\prime$-bosons couple to quarks and leptons not too weakly and if their mass
is not too large, they will be produced at the Tevatron and the LHC and easily
detected through their leptonic decay modes. The search 
for these particles occupies therefore an important place in the present
experimental program.

When studying the transverse-momentum ($p_T$) distribution of a $Z^\prime$-boson
with an invariant mass $M$, it is convenient to separate the large- and
small-$p_T$ regions. For the large values of $p_T$, the 
use of the fixed-order perturbation theory is fully justified, since the
perturbative series is controlled by a small expansion parameter,
$\alpha_s(M^2)$, but in the small-$p_T$ region, the coefficients of the
perturbative expansion are enhanced by powers of large logarithmic terms,
$\ln(M^2 / p_T^2)$. As a consequence, the convergence of the perturbative series
is spoiled as $p_T \to 0$. Furthermore, when the initial partons have just
enough energy to produce a $Z^\prime$-boson, the mismatch between virtual
corrections and phase-space suppressed real-gluon emission leads also to the
appearance of large logarithmic terms $[\ln(1-z)/(1-z)]_+$, where $z=M^2/s$, $s$
being the partonic centre-of-mass energy. However, the convolution of the
partonic cross section with the steeply falling parton distributions enhances
the threshold contributions even if the hadronic threshold is far from being
reached and large corrections are expected.

Accurate calculations of $p_T$ and invariant-mass distributions must
then include soft-gluon resummation in order to obtain reliable perturbative
predictions and properly take these logarithmic terms into account. We implement
the joint resummation formalism \cite{Bozzi:2007tea} at the next-to-leading
logarithmic (NLL) accuracy and consistently match the obtained result with the
pure perturbative result at the first order in $\as$ \footnote{Note that
next-to-next-to-leading order effects have recently been investigated as well
\cite{Coriano:2008wf}.},
which allows us to resum all the logarithms simultaneously. In addition, we
compare our predictions 
\cite{Fuks:2007gk} with those of the next-to-leading order (NLO) 
Monte Carlo generator MC@NLO \cite{Frixione:2002ik} and of the leading-order
(LO) Monte Carlo generator PYTHIA \cite{Sjostrand:2006za}, and study the impact
of the scale dependence.

%%%%%%%%%%%%%% Begin Section 2 %%%%%%%%%%%%%%%%%%%%%%%%%%%%%%%%%%%%%%%%%

\section{Theoretical framework} \label{sec:2}
\subsection{Considered $Z^\prime$ model} \label{subsec:2.1}

Ten-dimensional string theories with $E_8\times E_8$ gauge symmetry are
anomaly-free and contains chiral fermions as in the SM
\cite{Green:1984sg}. After compactification, this symmetry leads to an effective
 $E_6$ GUT group that can be broken further to 
\cite{Hewett:1988xc} 
\beq \label{eq:1}
 E_6\,\to\,SO(10) \times U(1)_\psi
    \,\to\,SU(5)  \times U(1)_\chi \times U(1)_\psi.~
\eeq While the $Z^\prime$-bosons related to the additional $U(1)_\psi$ and
$U(1)_\chi$ symmetries can in general mix, we consider in this
work only a TeV-scale $Z_\chi$-boson and assume the $Z_\psi$ to acquire its mass
at much higher scales, as it is naturally the case in the hierarchy of
symmetry breaking of Eq.\ (\ref{eq:1}).

\subsection{Joint resummation formalism at the next-to-leading logarithmic
level} \label{sec:2.2}

In Mellin $N$-space, the resummed hadronic cross section for the hard scattering
process 
\beq \vspace*{-2mm}
  h_a\, h_b \to  \gamma, Z, Z^\prime \to l^+ l^- (M, p_T) + X,~
\eeq where a lepton pair with invariant mass $M$ and transverse momentum
$p_T$ is produced, factorizes 
\beq  \vspace*{-2mm} 
  \frac{\d^2\sigma^{( \rm res )}}{\d M^2\, \d p^2_T}(N,b) =
    \sum_{a,b} f_{a/h_a}(N+1)\, f_{b/h_b}(N+1)\, {\cal H}_{ab}(N)\,
    \exp\{{\cal G}(N,b)\}.~  
\eeq $f_{a,b/ h_{a,b}}$ are the $N$-moments of the universal
distribution 
functions of partons $a,b$ inside the hadrons $h_{a,b}$, the impact-parameter
$b$ is the variable conjugate to $p_T$ through a Fourier transform and the
dependence on the renormalization and factorization scales $\mu_R$
and $\mu_F$ has been removed for brevity. The process-dependent function ${\cal
H}$ contains all the  terms due to hard virtual corrections and collinear
radiation while the process-independent Sudakov form factor ${\cal G}$ allows to
resum the soft-collinear radiation, embodying the all-order dependence on the
logarithms.

Once resummation has been achieved in $N$- and $b$-space, inverse transforms
have to be performed in order to get back to the physical space. Since the
resummed exponent contains singularities, the integration contours of the
inverse transforms must avoid hitting any of these poles. The $b$-integration is
performed by deforming the integration contour with a diversion into the complex
$b$-space \cite{Laenen:2000de}, while the inverse Mellin transform
is performed following a contour inspired by the Minimal
Prescription \cite{Catani:1996yz} and the Principal Value
Resummation \cite{Contopanagos:1993yq}.

In order to keep the full information contained in the fixed-order calculation
and to avoid possible double-counting of the logarithmically enhanced
contributions, a matching procedure of the NLL resummed cross section to the
fixed order result is performed through the formula 
\beq 
  \frac{\d^2\sigma}{\d  M^2\,\d p_T^2} = \frac{\d^2\sigma^{({\rm F.O.})}}{\d
  M^2\,\d p_T^2} + \oint_{C_N} \frac{\d N}{2\pi i}\, \tau^{-N}\!\! \int
  \frac{b\, \d b}{2} J_0(b\, p_T) \left[\frac{\d^2\sigma^{{\rm (res)}}}{\d
  M^2\,\d p_T^2}(N, b) - \frac{{\rm d}^2\sigma^{{\rm (exp)}}}{\d
  M^2\,\d p_T^2}(N, b) \right].~
\eeq  $J_0(x)$ is the $0^{{\rm th}}$-order Bessel function, $\d^2\sigma^{({\rm
F.O.})}$ is the fixed-order perturbative result, $\d^2\sigma^ {({\rm res})}$ is
the resummed cross section and $\d^2\sigma^{({\rm exp})}$ is its truncation to
$\mathcal{O}(\as)$.

%%%%%%%%%%%%%% Begin Section 3 %%%%%%%%%%%%%%%%%%%%%%%%%%%%%%%%%%%%%%%%%

\section{Numerical results} \label{sec:3}

\begin{figure}
\centering
  \psfig{figure=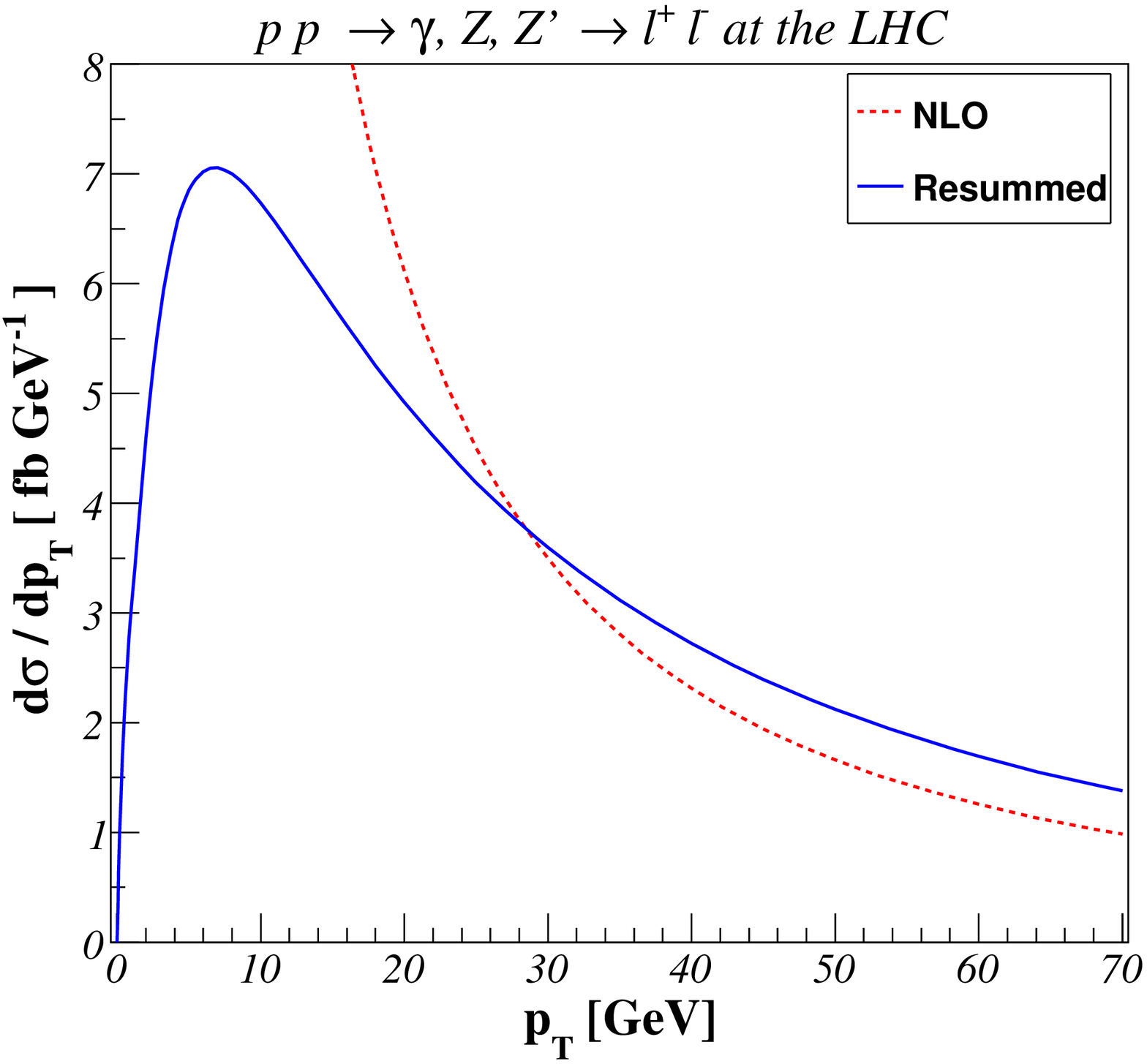,width=0.32\columnwidth}
  \psfig{figure=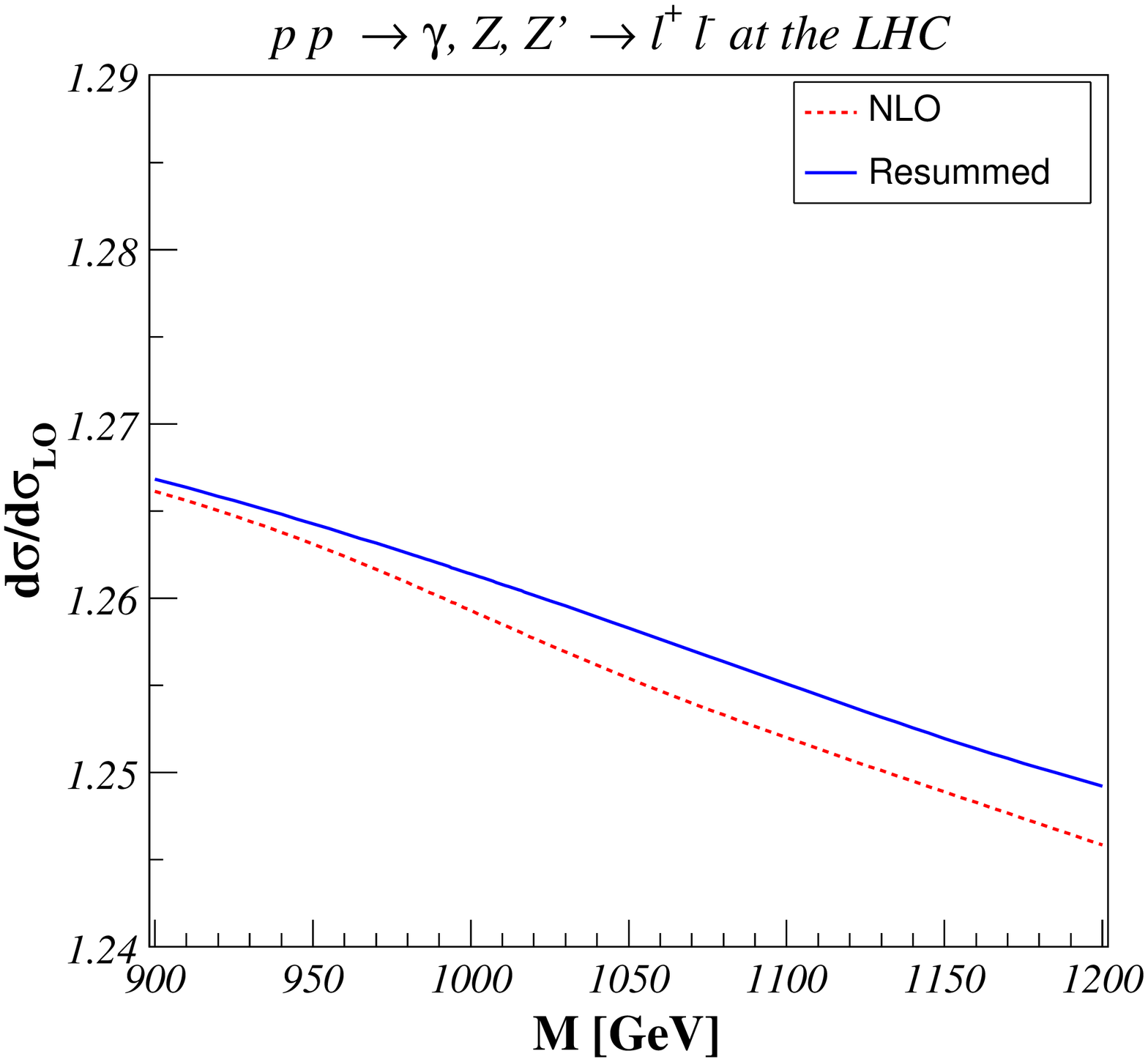,width=0.32\columnwidth}
  \psfig{figure=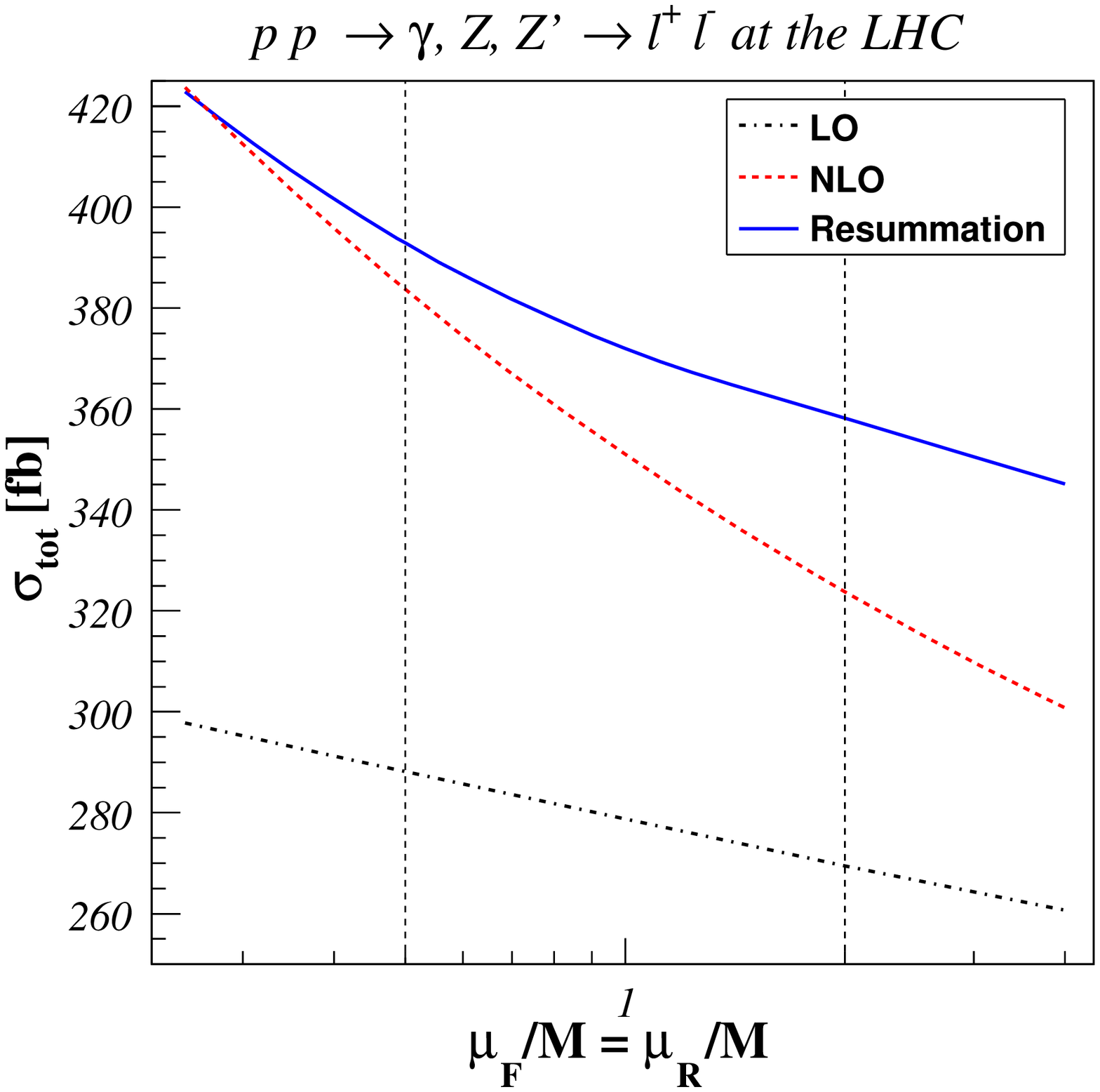 ,width=0.33\columnwidth}
\caption{\label{fig:01} Transverse-momentum spectra (left), mass spectra
 (centre) and dependence of the total cross section on the common
 factorization/renormalization scale (right)  in LO QCD (dot-dashed), NLO QCD
 (dashed) and after resummation (plain). The mass spectra have been normalized
 to the LO QCD prediction.} 
\end{figure}

In Fig.\ \ref{fig:01} (left), we show the $p_T$-spectrum (integrated over the
invariant mass in the range from 900 to 1200 GeV) for
$Z^\prime$-bosons with a mass of 1 TeV produced at the LHC. As expected, the true
NLO distribution (dashed) diverges as $p_T\to 0$ while resummation (plain) leads
to a smooth turnover with a maximum at around 8 GeV. As can be seen from Fig.\ 
\ref{fig:01} (centre), the NLO $K$-factor (dashed) is increased further by the
resummed contributions (plain), even if we are still relatively far from the
production threshold. The scale uncertainty of the total cross section
(integrated over all transverse momenta and over the invariant mass in the range
from 900 to 1200 GeV) is shown in Fig.\ \ref{fig:01} (right). The LO QCD
prediction (dot-dashed) does not give a reliable estimate of the theoretical
error, since the NLO cross section (dashed) is considerably larger. At NLO, the
factorization scale dependence is reduced as expected, but $\as(\mu_R)$ makes
its appearance, so that an additional renormalization scale dependence is
introduced. The total NLO scale dependence is reduced to 9\% (vertical lines)
once resummation is achieved (plain).

\begin{figure}
  \centering
  \psfig{figure=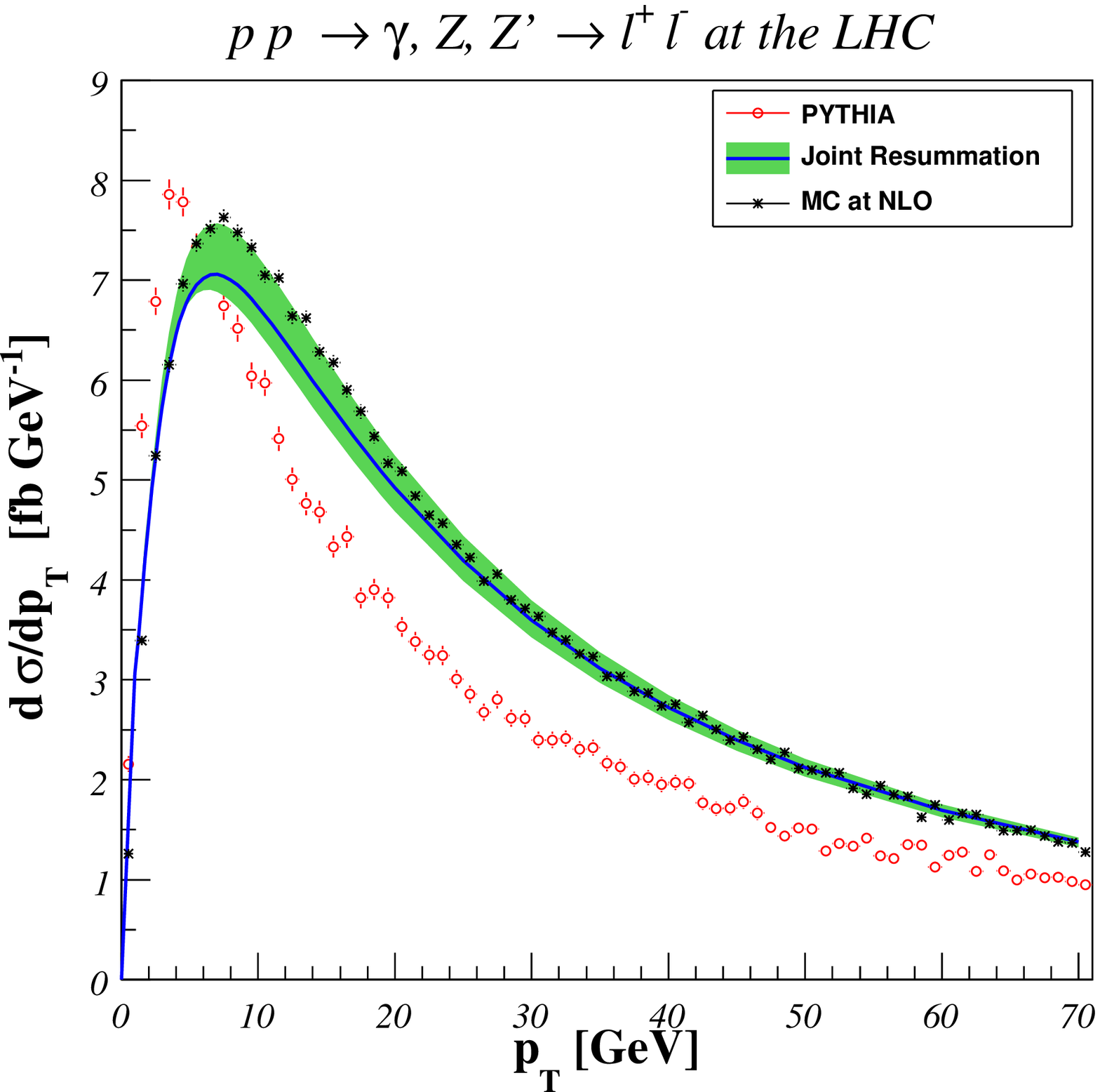,width=0.33\columnwidth}
  \psfig{figure=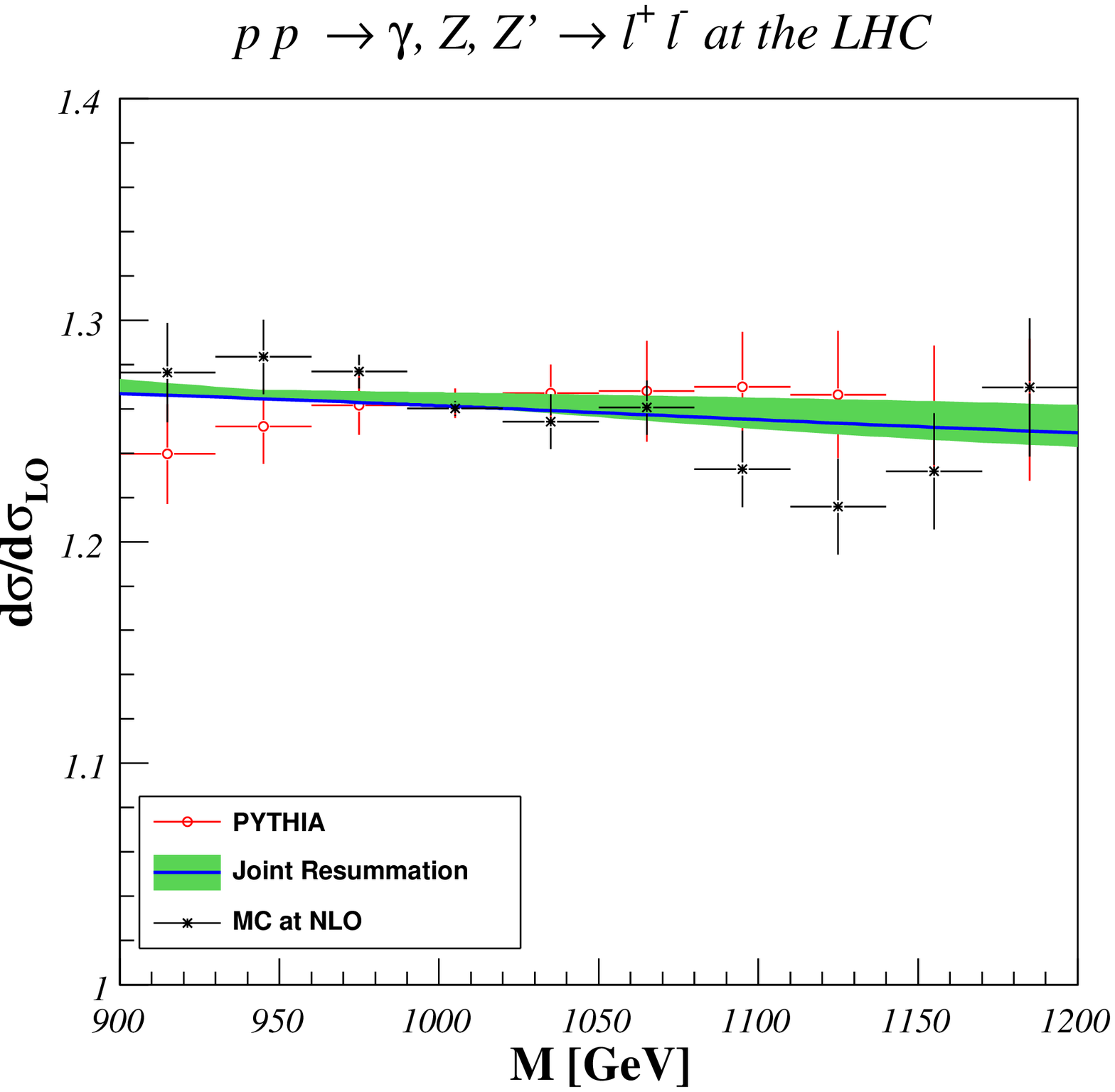,width=0.33\columnwidth}  
\caption{\label{fig:02} Mass (left) and transverse-momentum spectra (right)
 in PYTHIA (circles), in MC@NLO (stars), and after joint resummation
 (plain). The mass spectra have been normalized to the LO QCD
 prediction, and the renormalization and factorization scale uncertainties
 in the resummed predictions are indicated as shaded bands.}
\end{figure}

We confront in Fig.\ \ref{fig:02} the resummed results (plain) to the LO
predictions obtained with PYTHIA's ``power shower'' (circles) and to the NLO
predictions obtained with MC@NLO's parton shower (stars). The shaded band around
the resummed predictions corresponds to the theoretical uncertainty induced
through the simultaneous variation of the renormalization and factorization
scales by a factor of two around the central scale $M$. The $p_T$-spectra (left)
are, for all three calculations, no longer divergent. 
The PYTHIA prediction rises and falls rather steeply around its maximum at 3
GeV, whereas the MC@NLO and resummed predictions rise and fall more slowly
around the peak at 8 GeV. The agreement between MC@NLO and joint resummation is
impressive, in particular for a scale choice of $M/2$ (upper end of the shaded
band).  The correction factors for the mass spectra (right), which have been
normalized to the LO QCD prediction, show only a very weak mass dependence. We
have multiplied the PYTHIA mass spectrum by hand with a 
global $K$-factor of 1.26. Otherwise, within the statistical error bars, the
PYTHIA $K$-factor would just be unity, since the normalization of the total
cross section is not changed by the parton shower. Again, the MC@NLO $K$-factor
agrees almost perfectly with the one of resummation.

%%%%%%%%%%%%%% Begin Section 4 %%%%%%%%%%%%%%%%%%%%%%%%%%%%%%%%%%%%%%%%%

\section{Conclusions} \label{sec:4}

We have improved the theoretical predictions for the production
of extra neutral gauge bosons at hadron colliders, which are currently based
on the LO Monte Carlo generator PYTHIA, by implementing the $Z^\prime$-bosons in
the MC@NLO generator and by computing their differential and total cross
sections in the framework of joint resummation. MC@NLO and resummation were
found to be in excellent agreement for mass spectra and $p_T$ spectra, while the
PYTHIA predictions show significant shortcomings both in normalization and
shape. The theoretical uncertainties from scale variations 
were found to be under good control. The implementation of our improved
predictions in terms of the new MC@NLO generator or resummed $K$ factors in the
analysis chains of the Tevatron and LHC experiments should be straightforward
and lead to more precise determinations or limits of the $Z^\prime$-boson masses
and/or couplings.

\section*{Acknowledgments}
The author acknowledges the conference organizing committee for financial
support.


\section*{References}

\begin{thebibliography}{99}

\bibitem{Langacker:2008yv}
  P.~Langacker,
  %``The Physics of Heavy Z' Gauge Bosons,''
  arXiv:0801.1345 [hep-ph], and references therein.
  %%CITATION = ARXIV:0801.1345;%%
  
\bibitem{Bozzi:2007tea}
  G.~Bozzi, B.~Fuks and M.~Klasen,
  %``Joint resummation for slepton pair production at hadron colliders,''
  Nucl.\ Phys.\  B {\bf 794} (2008) 46.
  %[arXiv:0709.3057 [hep-ph]].
  %%CITATION = NUPHA,B794,46;%%

\bibitem{Coriano:2008wf}
  C.~Coriano, A.~E.~Faraggi and M.~Guzzi,
  %``Searching for Extra Z' from Strings and Other Models at the LHC with
  %Leptoproduction,''
  arXiv:0802.1792 [hep-ph].
  %%CITATION = ARXIV:0802.1792;%%  

\bibitem{Fuks:2007gk}
  B.~Fuks, M.~Klasen, F.~Ledroit, Q.~Li and J.~Morel,
  %``Precision predictions for Z'-production at the CERN LHC: QCD matrix
  %elements, parton showers, and joint resummation,''
  Nucl.\ Phys.\  B {\bf 797} (2008) 322.
  %[arXiv:0711.0749 [hep-ph]].
  %%CITATION = NUPHA,B797,322;%%

\bibitem{Frixione:2002ik}
  S.~Frixione and B.~R.~Webber,
  %``Matching NLO QCD computations and parton shower simulations,''
  JHEP {\bf 0206} (2002) 029.
  %[arXiv:hep-ph/0204244].
  %%CITATION = JHEPA,0206,029;%%

\bibitem{Sjostrand:2006za}
  T.~Sj\"ostrand, S.~Mrenna and P.~Skands,
  %``PYTHIA 6.4 physics and manual,''
  JHEP {\bf 0605} (2006) 026.
  %[arXiv:hep-ph/0603175].
  %%CITATION = JHEPA,0605,026;%%

\bibitem{Green:1984sg}
  M.~B.~Green and J.~H.~Schwarz,
  %``Anomaly Cancellation In Supersymmetric D=10 Gauge Theory And Superstring
  %Theory,''
  Phys.\ Lett.\  B {\bf 149} (1984) 117.
  %%CITATION = PHLTA,B149,117;%%

\bibitem{Hewett:1988xc}
  J.~L.~Hewett and T.~G.~Rizzo,
  %``Low-Energy Phenomenology of Superstring Inspired E(6) Models,''
  Phys.\ Rept.\  {\bf 183} (1989) 193.
  %%CITATION = PRPLC,183,193;%%

\bibitem{Laenen:2000de}
  E.~Laenen, G.~Sterman and W.~Vogelsang,
  %``Higher-order QCD corrections in prompt photon production,''
  Phys.\ Rev.\ Lett.\  {\bf 84} (2000) 4296.
  %[arXiv:hep-ph/0002078].
  %%CITATION = PRLTA,84,4296;%%

\bibitem{Catani:1996yz}
  S.~Catani, M.~L.~Mangano, P.~Nason and L.~Trentadue,
  %``The Resummation of Soft Gluon in Hadronic Collisions,''
  Nucl.\ Phys.\  B {\bf 478} (1996) 273.
  %[arXiv:hep-ph/9604351].
  %%CITATION = NUPHA,B478,273;%%

\bibitem{Contopanagos:1993yq}
  H.~Contopanagos and G.~Sterman,
  %``Principal value resummation,''
  Nucl.\ Phys.\  B {\bf 419} (1994) 77.
  %[arXiv:hep-ph/9310313].
  %%CITATION = NUPHA,B419,77;%%
  
\end{thebibliography}
\end{document}